\documentclass[useAMS,usenatbib]{mn2e}
\usepackage{epsfig}
\usepackage{amsmath}
\usepackage{amssymb}

\newcommand{\beqa}{\begin{eqnarray}}
\newcommand{\eeqa}{\end{eqnarray}}
\newcommand{\be}{\begin{equation}}
\newcommand{\beq}{\begin{equation}}
\newcommand{\ba}{\begin{eqnarray}}
\newcommand{\ee}{\end{equation}}
\newcommand{\eeq}{\end{equation}}
\newcommand{\ea}{\end{eqnarray}}
\newcommand{\Msun}{M_{\odot}\hspace{1mm}}
\newcommand{\msun}{$M_{\odot}\hspace{1mm}$}
\newcommand{\wmap}{{\it WMAP }}

\newcommand{\HI}{H{\scriptsize{\textrm{I}}}}
\newcommand{\HII}{H{\scriptsize{\textrm{II}}}}

\newcommand{\HeII}{He{\scriptsize{\textrm{II}}}}

\newcommand{\Lylim}{Lyman limit}
\newcommand{\fstar}{f^\star}
\newcommand{\fstaresc}{f^*_{{\mathrm{esc}}}}
\newcommand{\fesc}{f_{{\mathrm{esc}}}}
\newcommand{\zsix}{$z \sim 6$}
\newcommand{\omegab}{\Omega_{\mathrm{B}}}
\newcommand{\omegam}{\Omega_{\mathrm{M}}}
\newcommand{\omegal}{\Omega_{\mathrm{\Lambda}}}
\newcommand{\Lnu}{L_{{\scriptscriptstyle{\nu}}}}
\newcommand{\nul}{\nu_{\scriptscriptstyle{\mathrm{L}}}}

\newcommand{\flyescplus}{f^{\sss{1500}}_{\mathrm{esc}}}

\newcommand{\Lya}{Ly{$\alpha$}}

\newcommand{\Feff}{F_{\scriptscriptstyle\mathrm{eff}}}
\newcommand{\Fbareff}{\bar{F}_{\scriptscriptstyle{\rm eff}}}
\newcommand{\Fbareffobs}{\bar{F}^{\sss{\mathrm{obs}}}_{\scriptscriptstyle{\rm eff}}}

\newcommand{\lambdaMFP}{\lambda_{\scriptscriptstyle{\mathrm{mfp}}}}

\newcommand{\vectorx}{{{\mathbf{x}}}}

\newcommand{\jtw}{J_{\mathrm{\mathbf{\scriptstyle 21}}}}
\newcommand{\JHI}{J_{\scriptscriptstyle \mathrm H{\scriptsize{\textrm{I}}}}}
\newcommand{\nHI}{n_{\scriptscriptstyle \mathrm H{\scriptsize{\textrm{I}}}}}

\newcommand{\ts}{t_{\scriptscriptstyle{\mathrm s}}}
\newcommand{\tH}{t_{\scriptscriptstyle{\mathrm H}}}
\newcommand{\tdc}{t_{\scriptscriptstyle\mathrm{DC}}}

\newcommand{\sssrm}{\scriptscriptstyle\mathrm}
\newcommand{\scs}{\scriptstyle}

\newcommand{\sss}{\scriptscriptstyle}
\newcommand{\pfk}{P_{\scriptscriptstyle{\rm F}}(k)}

\def\lsim{~\rlap{$<$}{\lower 1.0ex\hbox{$\sim$}}}
\def\gsim{~\rlap{$>$}{\lower 1.0ex\hbox{$\sim$}}}

\title[Escape Fraction]{The escape fraction of ionizing photons from high redshift galaxies}

\author[Srbinovsky \& Wyithe]{J.A. Srbinovsky \& J.S.B. Wyithe\\
School of Physics, University of Melbourne, Parkville, Victoria, Australia 
Email: jsrbino@physics.unimelb.edu.au}

\date{Draft Version}
\pagerange{\pageref{firstpage}--\pageref{lastpage}}
\pubyear{2006}

\begin{document}
\label{firstpage}
\maketitle
\begin{abstract}
The fraction of ionizing photons which escape their host galaxy and so are able to ionize hydrogen in the inter-galactic medium (IGM) is a critical parameter in studies of the reionization era and early galaxy formation.
In this paper we combine observations of Ly$\alpha$ absorption towards high redshift quasars with the measured UV luminosity function of high redshift galaxies to constrain the escape fraction ($\fesc$) of ionizing photons from galaxies at $z\sim5.5-6$.
We employ an $N$-body simulation to describe the density and peculiar velocity fields and to identify virialized halos.
To model the ionizing background (IBG) we associate these halos with star-bursting galaxies using a semi-analytic prescription.
We extract ensembles of mock {\Lya} absorption spectra for a range of values of $\fesc$ assuming all resolved ($M\gtrsim10^9$~\msun) simulation halos are able to contribute to the IBG. 
The observed Ly$\alpha$ transmission constrains the escape fraction to lie in the range $\fesc\sim10-25\%$ (at $z\sim5.5-6$). 
Excluding halos with $M< 10^{10} M_{\odot}$ (as might be expected if galaxy formation is suppressed due to the reionization of the IGM) implies a larger escape fraction of $\fesc\sim20-45\%$. 
Using the numerical results to calibrate an analytic relation between the escape fraction and minimum galaxy halo mass we also extrapolate our results to a mass ($M\sim10^8$~\msun) corresponding to the hydrogen cooling threshold.
In this case we find $\fesc\sim5-10\%$, consistent with observed estimates at lower redshift.
We find that the escape fraction of high redshift galaxies must be greater than $5\%$ irrespepctive of galaxy mass.
Based on these results we use a semi-analytic description to model the reionization history of the IGM, assuming ionizing sources with escape fractions suggested by our numerical simulations. 
We find that the IBG observed at $z\sim5.5-6$ implies a sufficient number of ionizing photons to have reionized the Universe by \zsix.
However, if the minimum mass for star-formation were $\gtrsim10^9$~\msun, the IBG would be over-produced at $z\lesssim5$.
In summary, our results support a scenario in which the IGM was reionized by low mass galaxies. 
\end{abstract}
\noindent 

\section{INTRODUCTION}
\label{introduction_section}

The sources thought to have produced the UV radiation which reionized the hydrogen gas in the inter-galactic medium (IGM) are star-bursting galaxies and quasars [e.g. \citet{Barkana2001}]. 
However, the number  density of the quasar population is observed to decline exponentially with redshift at $z\gtrsim 2.5$, implying that galaxies probably contribute the bulk of UV photons which drive reionization at $z\gtrsim 6$ and also the evolution of the ionizing background post-overlap between $ z \sim 3-6$ \citep{Madau1999,Fan2002,Srbinovsky2007,Bolton2007a}.
The potential contribution of galaxies to the UV radiation field is a function of their number density, the star formation rate (SFR) within the galaxies and the spectral energy distribution (SED) of the stellar population.
However, the fraction of ionizing photons which escape their host galaxy into the IGM, is limited by intervening absorption in the inter-stellar medium (ISM).
The value of this escape fraction ($\fesc$) is therefore a critical parameter in studies of reionization, and remains relatively uncertain. 

Owing to its importance, constraints on the value of the escape fraction feature extensively in the literature.
However, the definition of $\fesc$ varies depending on the study.
A tendency exists in observational studies to quote values of $\fesc$ as the escape fraction of Lyman-limit photons, relative to the escape fraction at $1500\AA$ ($\flyescplus$).
In this paper, we define $\fesc$ as the absolute escape fraction of photons at the Lyman-limit, since this is the parameter of physical importance for the reionization history of hydrogen.
Where necessary, we convert values of the escape fraction from published studies to be consistent with this definition, assuming $\flyescplus=0.24$ and an extinction with a median value of E(B-V) $=0.15$ at $z \lesssim 3$ \citep{Siana2007}. 

Early theoretical studies predicted values for the escape fraction of between $\fesc \sim 3-15\%$ \citep{Dove1994, Dove2000,Ciardi2002, Fujita2003}, however \citet{Clarke2002} suggested that $\fesc$ could be much higher than this if the feedback dependent porosity of the ISM is considered.
At high redshift, \citet{Fujita2003} argue that increased supernovae activity create escape tunnels through the ISM for ionizing radiation, and estimate $\fesc>20\%$ by $z>5$. Similarly, \citet{Razoumov2006} find that $\fesc$ evolves from $\sim1-2\%$ at $z=2.39$ to $\sim6-10\%$ at $z=3.6$.  
In contrast, \citet{Wood2000} suggest that $\fesc$ decreases with redshift (becoming $\sim 1\%$ at $z\sim 10$) due to the increasing density of the galactic disk. 
At $z\sim20$, \citet{Whalen2004} and \citet{Alvarez2006}, have derived escape fractions as high as $\fesc\sim 95\%$, however in these cases the large value can be attributed to the assumed population-III stellar population.
Finally, \citet{Kitayama2004} studied the evolution of \HII~regions surrounding a central population-III star embedded in a galaxy and find that $\fesc$ increases with the mass of the star and inversely with the mass of the halo.

In the local Universe, observational estimates of $\fesc \sim 2-11\%$, have been derived\footnote{In this range for $\fesc$ we have excluded the finding of \citet{Hurwitz1997} for one of the four galaxies in their studied sample in which $\fesc \lesssim 57\%$.}
from measurements of emission line strengths relative to the Lyman-limit flux [\citet{Leitherer1995,Hurwitz1997,Heckman2001,Bergvall2006,Grimes2007}].
An alternative approach measures the flux in the Lyman-continuum relative to the flux at a more readily detectable frequency above the Lyman-limit (usually $1500\AA$).
Comparison is then made to the intrinic flux ratio predicted by an assumed  SED of galaxies, which can be characterized by the flux decrement across the Lyman break ($LB$).
Assuming $LB=3$, \citet{Deharveng2001} estimates values of $\fesc\lesssim 10\%$ in nearby galaxies, and \citet{Malkan2003} estimates values of $\fesc\lesssim 1-6\%$ from a sample of $11$ blue galaxies at $z\sim 1$.
At $z\sim3$, \citet{Steidel2001} compile a composite spectrum from $29$ Lyman-break galaxies and find an average escape fraction of $\fesc>10\%$, and from observations of 2 galaxies  \citet{Giallongo2002} determine values of $\fesc \lesssim 4\%$.
From a sample of 14 star-forming galaxies, \citet{Shapley2006} determines the escape fraction to be  $\fesc\sim 3\%$, however, in two of these galaxies detection of Lyman continuum flux suggests an escape fraction of the order of $10\%$.
A less stringent constraint is found by \citet{Inoue2005}, who find the escape fraction to be $\fesc \lesssim 20-40 \%$ at $z\sim 3$ using narrow band images of two galaxies.
Finally, the escape fraction has also been constrained by \citet{Schaye2006} who found $\fesc\sim10\%$ at $z\sim3$, by comparing the absorption systems near an ionizing source with those in the average IGM.

Alternatively, if an SED is available for the galaxy, then the 
the effects of intrinsic and intervening absorption can be modelled.
Using this method \citet{Fernandez2003} find $\fesc \leq 4\%$, by fitting to the observed SEDs of 27 galaxies at $1.9\leq z\leq 3.5$.
Similarly, at $z\sim1$, \citet{Siana2007} fits optical/near-infrared model SEDs to the observed spectra of 21 galaxies and finds the average, intrinsic Lyman-break factor to be $\sim 6$, from which they conclude the global escape fraction to be, $\fesc \lesssim 4\%$.
Recently, \citet{Chen2007} estimate $\fesc\sim2\%$ at $z\sim2$ from the observed distribution of neutral hydrogen column densities in the after-glow spectra of long duration GRBs.
This approach alleviates the dependence on largely unknown intrinsic spectral properties of the galaxy, however presents another issue of a possible GRB environment bias.
Finally, combining estimates of $\fesc$ from published studies with values derived from measurements of the ionizing background, \citet{Inoue2006} find that the escape fraction increases from $\fesc \sim 1-10\%$ in the redshift interval $1 \lesssim z \lesssim 4$. 

While there is as yet no consensus on the value of $\fesc$, the implications for the reionization history have been addressed by several authors.
Recently, detailed numerical simulations \citep{Gnedin2007a} have predicted a value for $\fesc$ of between $1$ and $3\%$, from halos of mass $M\gtrsim5\times10^{\scs 10}M_{\odot}$ within the redshift range $3<z<9$. 
In addition to a small escape fraction in massive galaxies \citet{Gnedin2007a} further predict that halos with $M\lesssim 5\times10^{\scs 10}M_{\odot}$ have an escape fraction which is negligibly small.
This very low efficiency of reionization would have profound effects on the reionization history.
Indeed, \citet{Gnedin2007b} argues that the small escape fraction for lower mass galaxies implies that the observed galaxy population at $z\sim6$ could not have produced enough UV photons to reionize the IGM.
This suggests that the process of reionization takes place in a ``photon starved" environment.
This conclusion has been previously reached by \citet{Bolton2007a} who studied the ionizing background (IBG) based on the {\Lya} absorption properties towards high redshift quasars. 
However, \citet{Bolton2007a} inferred a higher a value for the escape fraction of $\fesc \sim 20-30\%$ (at \zsix).

In this paper we employ an approach that is similar to \citet{Bolton2007a}, and constrain the escape fraction of ionizing photons (at $z\sim5.5-6$) by combining the observed luminosity function (LF) of galaxies with the observed opacity of the high redshift IGM. 
However, with respect to the study of the escape fraction our modelling includes several improvements.
We employ an $N$-body simulation to describe the density and peculiar velocity fields and to identify virialized halos.
We determine the UV luminosity of each halo using a semi-analytic description and constrain model parameters by fitting to the observed galaxy LF.
We then generate an inhomogeneous ionizing background using these discrete galaxies, which depends on the escape fraction of ionizing radiation.
For a range of escape fractions we extract an ensemble of mock {\Lya} absorption spectra along 3072 sight-lines through our simulation box and determine the value of $\fesc$ which best models the observed {\Lya} transmission.
We note that in our model the escape fraction is not explicitly dependent on galaxy mass or angular direction. 

The paper is organized as follows. 
In \S~\ref{model_section} we outline our models for the IBG due to discrete galaxies.
We also describe our computation of the {\Lya} photon transmission.
In \S~\ref{Gamma_and_escape_fraction_section} we show our results for $\fesc$ at several redshifts assuming MFPs which are consistent with the observationally derived values from \citet{Fan2006a}.
In \S~\ref{galaxy_suppression_section} we repeat our analysis to constrain $\fesc$ assuming the suppression of galaxy formation in low mass halos ($M<10^{10}M_{\odot}$), while in  \S~\ref{reionization_massive_galaxies_section} we exclude halos with $M<5\times10^{10}M_{\odot}$.
In \S~\ref{extrapolation_section} we extrapolate our numerical results to other minmum galaxy masses.
In \S~\ref{small_escape_fraction_models_section} we discuss the implications of our results using a semi-analytic model of reionization history. 
Finally we make some concluding remarks in \S~\ref{conclusions_section}.
Throughout the paper we assume a $\Lambda$CDM cosmology with parameters ($\sigma_{\scs 8}, h, \omegab, \omegam, \omegal$)=$(0.8,0.7,0.04,0.3,0.7)$. 
All distances are in co-moving units unless otherwise stated.


\section{Modelling {\Lya} transmission and a fluctuating ionizing background}
\label{model_section}

A brief summary of our approach to constrain the escape fraction is outlined below, followed by several subsections describing our model in more detail.
We employ an $N$-body simulation to compute the density and peculiar velocity fields and also to locate dark matter (DM) halos, which we then associate with galaxies.
To a fraction of these galaxies which are considered to be actively star-bursting, we assign a luminosity (at $\lambda=1350\AA$) based on models for the SFR and the intrinsic SED.
We constrain parameters in these models by measuring the LF from the simulation and fitting to the observed LF.
Assuming the IBG to be dominated by galaxies (see \S~\ref{introduction_section}), we next compute the luminosity of these actively UV luminous galaxies at the Lyman-limit, and using a simple model generate a field of ionizing radiation which reflects the distribution of these discrete sources.
We are then able to compute an ensemble of \Lya~forest spectra along lines-of-sight through our simulation volume, from which we can measure the mean transmission of the IGM.
Our model for describing the IBG contains the parameter $\fesc$, which we adjust so that the mean transmission measured from our model matches the observed mean transmission.

\subsection{The $N$-body simulation}
\label{The_N-body_section}

The simulation followed $1024^3$ DM particles in a cubic volume of side-length, $L=65.6$~Mpc/$h$.   
Snap-shots of the density and peculiar velocity fields were taken periodically in redshift space with bound halos identified using a {friends-of-friends} algorithm.
The simulation accurately resolves halos with mass, $M>2\times10^9 M_{\odot}$.
We place the DM halos and the density and velocity fields onto a $512^3$ grid.
In this paper we consider redshift snap-shots of the density and peculiar velocity fields at redshifts $z=5.5, 5.7, 6.0$.
The numerical simulation assumed $\sigma_{\scs 8}=0.9$ which we translate to a lower value of $\sigma_{\scs 8}$ following \citet{McQuinn2007}.
Further details of the numerical simulation can be found in \citet{Zahn2007,McQuinn2007} and \citet{Lidz2007}.

Baryonic gas pressure is not included in our (DM only) $N$-body simulation. 
However Hydro-Particle-Mesh (HPM) simulations can be used to approximate the effect of gas pressure.
\citet{Gnedin1998} show that HPM simulations of the IGM are capable of reproducing measurable quantities (such as absorption spectra) with a precision comparable to full hydro-dynamic simulations.
Though less reliable than HPM simulations, \citet{Gnedin1998} also show that the effect of pressure can be incorporated in $N$-body simulations by smoothing the linear density field on a suitable filtering scale ($k_{\scs f}$).
\citet{Lidz2006} measured the power spectrum of {\Lya} transmission fluctuations $\left[\pfk\right]$, using HPM simulations to describe the density and velocity fields.
To determine the correct filtering scale for our $N$-body simulation we therefore measure $\pfk$ following smoothing of the density field on a range of filtering scales.
We find the best agreement with \citet{Lidz2006} is obtained for a filtering scale of $k_{\scs f} = 20~{\mathrm{Mpc}}^{\scs -1}$, and adopt this value throughout the paper. 


\subsection{The ionizing background}
\label{ionizing_background_subsection}

The fraction of galaxies identified in our simulation which are actively star-bursting at a particular redshift, can be described by the duty-cycle of the star-burst ($\tdc=\ts/\tH$, where $\tH$ is the Hubble time at $z$).
To this fraction of galaxies we assign a luminosity (at $\lambda=1350\AA$) by first estimating the ${\mathrm{SFR}}$ which [following \citet{Loeb2005}] is dependent on the mass of the halo ($M$), the star-formation efficiency ($\fstar$) and again the duration of the star-burst ($\ts$).

\begin{align}
\label{SFR_equation}
{\mathrm{SFR}}=&0.17\left( \frac{M}{10^9~M_{\odot}}\right)
\left( \frac{\ts}{10^8~{\mathrm yr}}\right)^{-1} \notag \\
\times &\left( \frac{f^*}{0.1}\right)
\left( \frac{\omegab / \omegam}{0.17}\right)
~M_{\odot}~{\mathrm yr^{-1}} . 
\end{align}

\noindent
The specific luminosity at wavelength $\lambda$ of each halo can then be computed from

\begin{equation}
\label{luminosity_equation}
{\Lnu}(\lambda)=j_{\scriptscriptstyle\mathrm{L}}(\lambda)~\frac{\mathrm{SFR}}{1~M_{\odot}~{\mathrm yr^{-1}}} {\mathrm{ergs~s^{-1}Hz^{-1}}}  ,
\end{equation}

\noindent
where $j_{\scriptscriptstyle\mathrm{L}}(\lambda)$ is the specific rest frame luminosity, per Hz, per SFR (solar-masses per year).
We compute $j_{\scriptscriptstyle\mathrm{L}}(\lambda)$ using a model for SED of star-bursting galaxies which is presented in \citet{Leitherer1999}.
We assume metal enriched stars of $0.05$ solar metallicity, a \citet{Scalo1998} IMF (for  $1-100\Msun$) and continuous star-formation over a star-burst lifetime of 100 Myr.
This low solar metallicity is reasonable at high redshift and this star-burst lifetime is consistent with the duty-cycle  which is constrained below.
The relative flux density between $1500\AA$ and $900\AA$ for this SED is given by $LB\sim 3$.

\begin{figure}
\vbox{\centerline{
\epsfig{file=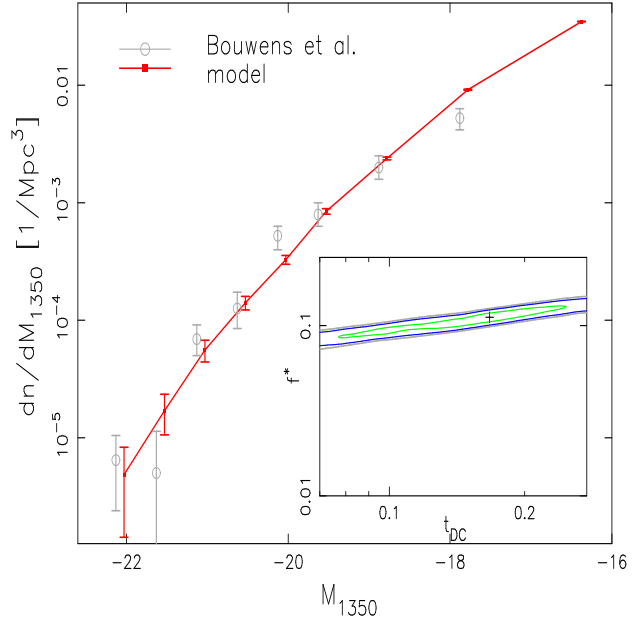,width=8.2truecm}}
\caption{\label{bestfitLF_NoS_figure}
{
The best fit model LF with parameter values of ${\tdc}=0.16$~(${\ts}=2.2\times10^8$~yr) and $f^*=0.11$, as constrained by a maximum likelihood analysis.
Inset is is a plot of the likelihood distribution in which the contours represent the $1,2,3-\sigma$ confidence levels with a cross to mark the point of maximum likelihood. 
For display purposes a line joins the model points (solid). 
The error bars on the model reflect the number of galaxies in each luminosity bin as measured from our simulation.
The open circles represent the data from the observed UV galaxy LF \citep{Bouwens2007}. 
The model points are shifted slightly to the right of the observational data for clarity only. 
}}}
\end{figure}

As our aim is to reproduce the observed LF  we correct this intrinsic source luminosity by including extinction by dust within the galaxy. 
We adopt a factor of $1.5$ for the ratio of the intrinsic to the observed luminosity (at $z\sim6 $) following \citet{Bouwens2007} [and references therein].
For $10,000$ combinations of ${\ts}$ and ${\fstar}$ we compute the LF in units of the number of galaxies per AB-magnitude per co-moving volume ($dn/dM_{1350}$) and constrain these parameters using a maximum likelihood technique, fitting our model LF to the observed LF of \citet{Bouwens2007} at $z=6$.
We assume flat prior probability distributions for the parameters $\tdc$ and $\fstar$.
In Figure~{\ref{bestfitLF_NoS_figure}} we show the best fit LF. 
For display purposes a line joins the model points (solid) which are the centers of the constructed luminosity bins.
At the faint end of the LF we have constructed an additional bin as within this star-formation model, the simulation resolves galaxies down to a lower luminosity than are currently observed.
The error bars on the model reflect the number of galaxies in each luminosity bin.
The open circles represent the data from the observed UV galaxy LF \citep{Bouwens2007}. 
The model points are shifted slightly to the right of the observational data for clarity. 
Inset in Figure~{\ref{bestfitLF_NoS_figure}} is a plot of contours representing the $1,2,3-\sigma$ confidence levels on the parameters $\fstar$ and $\tdc$.
A cross marks the point of maximum likelihood, with values of ${\tdc}=0.16~({\ts}=2.2\times10^8$~yr) and $f^*=0.11$. 

In this model the minimum mass halo ($M\sim2\times10^9 \Msun$) which contributes to the IBG corresponds to a specific luminosity of $L\sim 8.9\times10^{26}$ ergs/sec/Hz (at 1350~\AA) and a magnitude of $M_{\scs{1350}}\sim-15.8$.
The largest halo identified in our simulation has a mass of $M\sim3\times10^{12} \Msun$, which given the constrained parameters corresponds to a specific luminosity of $L\sim 1.2\times10^{30}$ ergs/sec/Hz and a magnitude of $M_{\scs{1350}}\sim-23.8$.
However at $z=6$, only $\sim 13\%$ of identified halos have masses greater than $1\times10^{10}\Msun$ and only $\sim 1\%$ have masses greater than $5\times 10^{10} \Msun$.
The largest actively star-bursting halo selected in our simulation has a mass of $M\sim1\times10^{12} \Msun$, which corresponds to a specific luminosity of $L\sim 4.4\times10^{29}$ ergs/sec/Hz and a magnitude of $M_{\scs{1350}}\sim-22.5$.
We note that in our model the minimum luminosity halo that contributes to the UV background is equivalent to the definition of the minimum luminosity cut off in the integration of the LF which is used to derive the global emissivity due to galaxies in other studies [e.g. \citet{Bolton2007a}].

Using the constrained values of $\fstar$ and $\tdc$, we re-evaluate the luminosity of each star-bursting galaxy at the Lyman-limit ($912\AA$).
We then re-position the UV luminous halos at the center of the simulation cell in which they reside and construct a luminosity grid [$L(\vectorx)$] as a function of position, which matches the dimensions of our density and velocity grids from the simulations.
As the grid is relatively fine ($65.6/512$~Mpc/$h$) the effect of this re-positioning on the resulting IBG is negligible in comparison to other uncertainties in our model (such as the assumed MFP).
More significantly however, the sources remain correlated with the over-density field, which is ultimately the desired effect here.
Very occasionally more than one halo resides within the boundaries of a cell.
In this case we consider a single more massive halo at the center of the cell.

Given the positions and luminosities of the galaxies we are able to compute the flux at every position on our grid using a procedure which is similar to that in \citet{Bolton2005,Bolton2007a,McDonald2005}.
We first compute the co-moving energy density ($\frac{dE}{dV}$) at the Lyman-limit through the Fourier convolution

\beq
\label{dEdV_conv_equation}
  \frac{dE}{dV}(\vectorx') = \frac{1}{V_{\rm cell}} \int_{-\infty}^{\infty} d\vectorx~ L(\vectorx) G(\vectorx'-\vectorx)~~,
\eeq

\noindent
which conveniently accounts for flux contributions from galaxies across the periodic boundaries in our simulation box.
In this equation $V_{\rm cell}$ is the co-moving volume of a grid cell and the attenuation of photons as they propagate from sources is described by the filter function $G({\vectorx'-\vectorx})$, which is defined as  
\begin{equation}
\label{filter_equation}
G({\vectorx'-\vectorx}) = \frac{e^{-\frac{|\vectorx'-\vectorx|}{\lambdaMFP}}}{4\pi (\vectorx'-\vectorx)^2c}~~,
\end{equation}

\noindent
where ${\lambdaMFP}$ is the MFP of ionizing photons, $\vectorx'$ is the position at which the energy density is to be evaluated, $\vectorx$ is the position of the source, and $c$ is the speed of light.
We then compute the flux in physical units of ${\jtw}$ (ergs/Hz/s/cm$^{\scs{2}}$/sr) from the co-moving energy density.
\beq
\label{j21_equation}
  \jtw(\vectorx') =\frac{\fesc (1+z)^2} {10^{\scs -21}}\frac{c}{4\pi} \frac{dE}{dV}(\vectorx')~~,
\eeq
Here, the (free) parameter $\fesc$ represents the fraction of ionizing photons produced during star-formation which escape the host galaxy.
An important simplification in the construction of our IBG is the assumption of a global, isotropic escape fraction and a universal MFP at each redshift. 

At $z=5.5, 5.7, 6.0$,  we take values of the MFP estimated by \citet{Fan2006a}, which assume a frequency averaged ionization cross-section, $\langle\sigma_{\scs \nu} \rangle$, corresponding to an IBG spectrum defined by $J_{\scs \nu} \propto \nu^{\scs -5}$.
We re-compute $\langle\sigma_{\scs \nu} \rangle$ using the spectrum appropriate for our model and adjust the values for $\lambdaMFP$ accordingly.
At the above redshifts we consider the MFPs, $\lambdaMFP=28.7, 18.9, 4.2$ Mpc/{\it h} respectively. 
Although these MFPs are derived from observational data, we note that they are model dependent and assume a uniform IBG.
We discuss the uncertainty in the MFP further in \S~\ref{escape_fraction_subsection}.

\subsection{{\Lya} absorption and the ionizing background}
\label{Lya_absorption_subsection}

Our simulation specifies (as a function of position) the density and the peculiar velocity of the gas as well as the ionizing radiation field.
We construct mock {\Lya} forest spectra as described in \citet{Hui1997} and briefly outline only the main points here, referring the reader to the original paper for details.
Along a given sight-line, over a distance between $\vectorx_{\sss A}$and $\vectorx_{\sss B}$, \citet{Hui1997} compute the optical depth as a function of observed frequency [{$\tau(\nu_o)$}] as 

\begin{equation}
\label{Hui_tau_equation}
   \tau(\nu_o)= \int_{\vectorx_{\sss A}}^{\vectorx_{\sss B}} \frac{dx}{1+\bar{z}}~ {\nHI}(\vectorx) {\sigma_{\sss \alpha}}~~.
\end{equation}
\noindent
Here ${\bar{z}}$ is the redshift at the center of the 1-D spatial interval $\vectorx_{\sss B}-\vectorx_{\sss A}$ in which $dx$ is the distance between cells, 
$\sigma_{\sss \alpha}$ is the {\Lya} cross-section and ${\nHI}(\vectorx)$ is the number density of neutral hydrogen.
We consider sight-lines parallel to the edges of the simulation box in which the perpendicular, central plane of the box corresponds to ${\bar{z}}$, which is also the redshift of our simulation snap-shot.

The {\Lya} cross-section is expressed as a function of the velocities $u_0$ and $u$ which can be interpreted in terms of the observed, the intrinsic and the resonant {\Lya} frequencies 

\begin{equation}
\label{Lya_crosssection_equation}
  \sigma_{\sss \alpha}=\sigma_{\sss{\alpha,0}} \frac{c}{b\sqrt{\pi}} e^{- \frac{(u-u_0)^{\sss 2}}{b^{\sss 2}} } ,
\end{equation}
\noindent
where $\sigma_{\sss{\alpha,0}}=4.5\times10^{\sss{-18}}$cm$^{\sss 2}$ and $c$ is the speed of light. 
The parameter $b=\left( 2k_{\sss B} T /m_{\sss p}\right)^{\sss{1/2}}$ is a function of the gas temperature ($T$) which accounts for the thermal broadening of the line profile. 
Here $k_{\sss B}$ is the Boltzmann constant and $m_{\sss p}$ is the mass of a proton.
\citet{Hui1997} find that the IGM gas temperature can be described locally by $T=T_{\scs 0}(1+\delta_b)^{\scs \gamma-1}$, with values of $1.2<\gamma<1.7$.
However, \citet{Lai2006} find the impact of temperature fluctuations on \Lya~forest spectra is small, therefore we assume $T_{\scs 0}=1.7\times10^{\scs 4}K$ and an isothermal IGM ($\gamma=1$).
We note that the uncertainty introduced by neglecting temperature fluctuations is absorbed into the uncertainty of our estimated value for the required ionization rate.
Along a sight-line, $u_0$ is determined in each pixel by the redshift relative to $\bar z$ and the corresponding velocity acquired due to the Hubble flow.
The parameter $u$ is determined by the peculiar velocity and the redshifted velocity of the {\Lya} resonance.

As a function of position we compute the proper number density of neutral hydrogen as
\beq 
\label{neutral_density_eq}
  {\nHI}(\vectorx)=f_{\nHI}(\vectorx) n_{\mathrm H}(\vectorx)~~,
\eeq
where $f_{\nHI}(\vectorx)$ is the neutral fraction and the density of hydrogen is given by $n_{\mathrm H}(\vectorx)= \bar{n}_{\mathrm H} \Delta(\vectorx)$.
The relative over-density [$\Delta(\vectorx)\equiv\rho(\vectorx)/\bar{\rho}$] is obtained from the $N$-body simulation and $\bar{n}_{\mathrm H}$ is the spatially averaged number density of hydrogen.
We compute the neutral fraction (following Hui et al.~1997) as
\beqa
\label{neutral_fraction}
  f_{\nHI}(\vectorx) &\sim& 1.6\times10^{-6}
    \left( \frac{T}{10^4~\rm{K}}    \right)^{-0.7}
    \left( \frac{\omegab h^2}{0.0125}    \right) \\ \notag
    &\times& \left( \frac{\JHI(\vectorx)}{0.5}    \right)^{-1}
    \Delta(\vectorx)\left( \frac{1+\bar{z}}{4}    \right)^{3}~~.
\eeqa

\noindent
In equation~(\ref{neutral_fraction}) the quantity $\JHI(\vectorx)$ is the average flux of ionizing photons in the IBG at wavelengths below the Lyman-limit given by 
\beq
\label{JHI_eq}
\JHI(\vectorx) = 
\frac  {\int_{\nul}^{\infty}  4\pi  J_{\nu} \frac {\sigma_{\nu}}   {h_{\mathrm{p}}\nu}  d\nu      }
  {  \int_{\nul}^{\infty}     4\pi        \frac {\sigma_{\nu}}   {h_{\mathrm{p}}\nu} d\nu      } ~~,
\eeq
where $J_\nu$ is the specific intensity as a function of frequency and $\sigma_\nu$ is the \HI~ photo-ionization cross section. 
For the spectrum of star-burst galaxies employed in this paper (\S~\ref{ionizing_background_subsection}), the ionizing flux can be expressed relative to the specific intensity at the Lyman-limit as ${\JHI}=0 .7{\jtw}$ [which we computed in equation~(\ref{j21_equation})].
We note that the numerator in equation~(\ref{JHI_eq}) is just the ionization rate [equation~(\ref{Gamma_equation})], highlighting the proportionality between the ionizing flux ($\JHI$), the ionization rate ($\Gamma_{-12}$) and the flux at the Lyman-limit ($\jtw$) via the SED.

We note that as it enters into the calculation of the optical depth [equations~(\ref{Hui_tau_equation}-\ref{JHI_eq})], $\fesc$ strictly implies the escape fraction of all ionizing photons.
Recent calculations of the frequency dependence of the escape fraction \citep{Gnedin2007a} show that $\fesc$ increases by a factor of (less than) $2$ between the \HI~($13.6$ eV) and \HeII~($54$ eV) ionization thresholds.
This suggests that interpreting $\fesc$ as the escape fraction at the Lyman-limit, leads to a small over-estimate of the quantity which is constrained via observations of galaxies as summarized in \S~\ref{introduction_section}.
However, the escape fraction at frequencies above the Lyman-limit are weighted by a factor proportional to the adopted SED.
Therefore, we expect the difference between $\fesc$ at the Lyman-limit and the escape fraction which is explicitly constrained by our model to be small in comparison to other uncertainties in our model.

Finally, the transmission at a particular frequency can be related to the optical depth by $F(\nu_o)={\mathrm{exp}}{\left[-\tau\left({\nu_o}\right)\right]}$.
Using the model for computing the optical depth presented above, we can construct \Lya-forest spectra by computing the transmission over a range of ($\nu_o$) frequencies .

\subsection{The observed transmissivity of the IGM}
\label{observed_transmissivity_section}

For a sample of 19 \Lya~absorption spectra observed towards high redshift quasars, \citet{Fan2006a} measure the average (or {\it{effective}}) transmission over a wavelength range in the \Lya~forest which corresponds to a redshift interval of $\Delta z=0.15$. 
From these transmission measurements along the available sight-lines, \citet{Fan2006a} compute the mean effective transmission ($\Fbareffobs$) in redshift slices (of thickness 0.2) of the IGM, which are centered on slightly different redshifts to the redshifts of our simulations ($\bar{z}$).
We therefore estimate values of ${\Fbareffobs}$ at $\bar{z}=5.5,5.7,6.0$ directly from the measurements of $\Feff$ from \citet{Fan2006a} (their Table~$2$) within redshift bins of 0.2 centered on each $\bar{z}$.
We compute the mean and the variance in this sample assuming a Gaussian distribution for uncertainties on individual transmission measurements.
At $z=5.5,5.7,6.0$ the observed mean effective transmission values (and $1-\sigma$ scatter) are ${\Fbareffobs}=0.079\pm0.035, 0.047\pm0.031, 0.006\pm0.005$.

We generate an ensemble of $3072$ absorption spectra for evenly spaced sight-lines throughout our simulation box.
Along each sight-line we then compute the effective transmission (${\Feff}$) over stretches of the forest which enables direct comparison with the observational data.
From the ensemble of ${\Feff}$ we calculate the mean effective transmission, ${\Fbareff}$.
Finally, we rescale the optical depth in each pixel (via $\fesc$) so that our simulated value of $\Fbareff$ matches the observed value.


\section{The Ionization Rate and the Escape Fraction}
\label{Gamma_and_escape_fraction_section}

In this section we present our estimates of the {\HI} ionization rate $(\Gamma)$ and the escape fraction of ionizing photons ($\fesc$), which we have constrained by adjusting the level of the IBG in our model so as to reproduce the transmission measurements (${\Fbareffobs}$) inferred from the spectra of high redshift quasars \citep{Fan2006a}. 

\subsection{Constraints on the ionization rate}
\label{Gamma_subsection}

\begin{figure}
\vbox{\centerline{
\epsfig{file=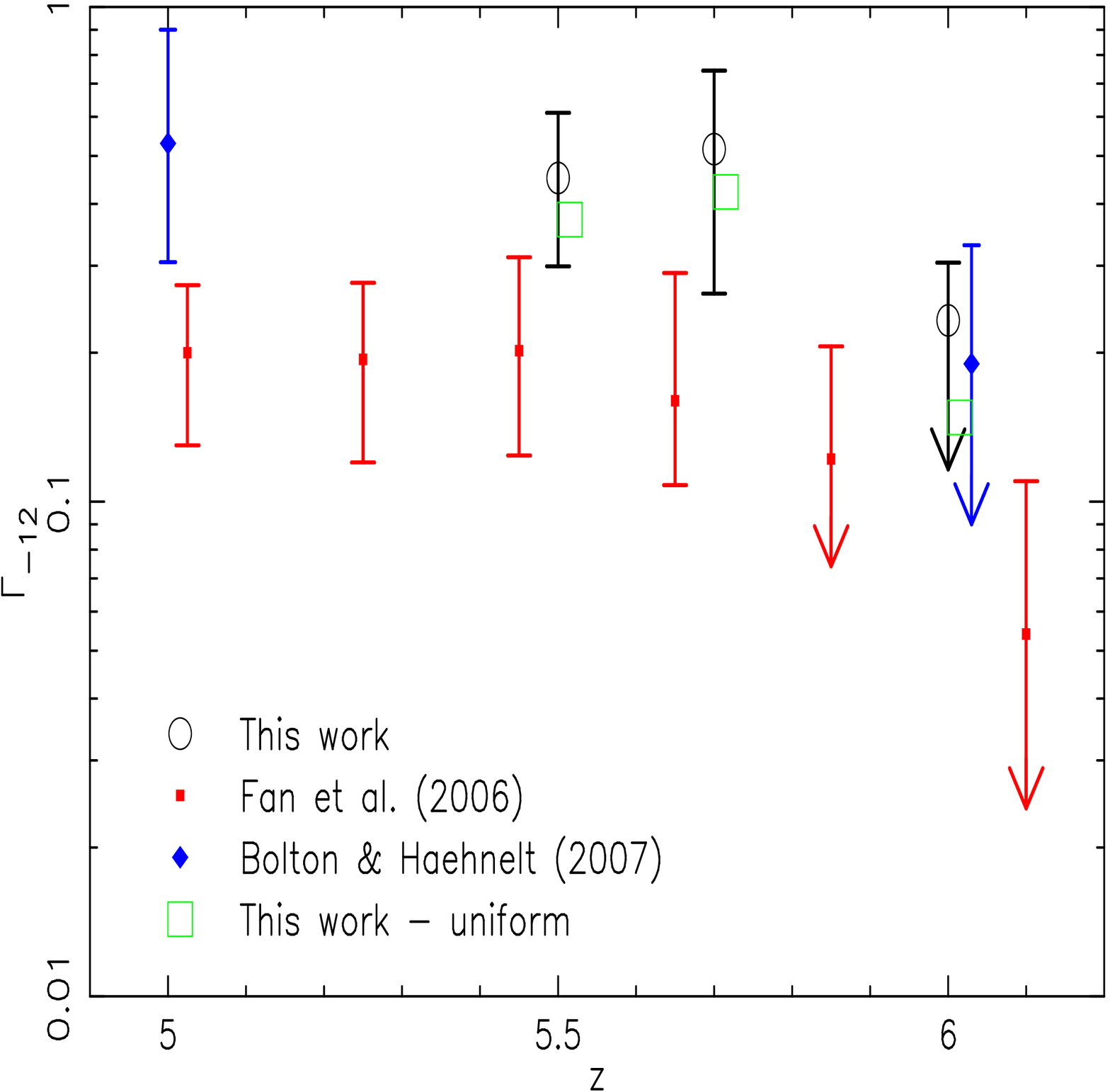,width=8.25truecm,height=6.5truecm}}
  \caption{\label{Gamma12_figure} {
Our estimates of the {\HI} ionization rates at $z=5.5,5.7,6.0$ assuming MFPs $\lambdaMFP=28.7, 18.9, 4.2$~Mpc/$h$ and mean transmission values of ${\Fbareff}=0.079, 0.047, 0.006$, respectively (open circles).
For comparison we show estimates of $\Gamma_{-12}$ from \citet{Bolton2007a} (diamonds) and \citet{Fan2006a} (points) which are also based on IGM {\Lya} transmission.
We also show our constrained values of $\Gamma_{-12}$ assuming uniform IBGs at each redshift (open squares).
} }}
\end{figure}

The {\HI} ionization rate is computed from the flux in the IBG ($J_{\nu}$) at all frequencies  above the {\Lylim} ({$\nul$}) by

\begin{equation}
\label{Gamma_equation}
\Gamma= 4\pi \int_{\nul}^{\infty} \frac{J_{\nu}}{h_{\mathrm{p}}\nu} \sigma_{\nu} d\nu~~,
\end{equation}

\noindent 
where $\sigma_{\nu}$ is the {\HI} photo-ionization cross-section, and $h_{\mathrm{p}}$ is Planck's constant.
For our spectrum of an IBG due to star-burst galaxies this relationship can be expressed as $\Gamma_{-12}= 2.8{\jtw}$,
where $\Gamma_{-12}$ is the photo-ionization rate in units of $10^{-12}$s${^{-1}}$.
To find the ionization rate in the IGM from the \Lya~forest spectra we trial values of $\fesc$, which is the free parameter we use to scale the intensity of the IBG.
For each $\fesc$, we use equation~(\ref{j21_equation}) to find $\jtw$, and hence the neutral fraction via equation~(\ref{neutral_fraction}). 
We compute \Lya~forest spectra from the absorption at a range of frequencies computed via equation~({\ref{Hui_tau_equation}}).
A value for the effective optical depth is computed from the \Lya~forest spectra along many sight-lines for a given $\fesc$.
The value of $\fesc$ is then adjusted until the simulation reproduces the observed, mean \Lya~ absorption.

In Figure~\ref{Gamma12_figure} we show the mean ionization rates constrained using our model for both a non-uniform background (open-circles) and a uniform background.
We note that in our model the ionization rate is spatially non-uniform as it follows the non-uniform IBG and so our estimates of the mean ionization rate are explicitly $\langle\Gamma_{-12}(\vectorx)\rangle$.
We compare our results to the ionization rates predicted by \citet{Bolton2007a} (diamonds) and \citet{Fan2006a} (points).
The $1-\sigma$ error in the ionization rate reflects the $1-\sigma$ uncertainty in the observed transmission at each redshift.
Our results are in excellent agreement with \citet{Bolton2007a} and a factor of $\sim 2$ greater than those of \citet{Fan2006a}. 
We refer the reader to the discussions in \citet{Bolton2005} and \citet{Bolton2007a} regarding the likely under-estimation of the ionization rates predicted by the fluctuating Gunn-Peterson approximation model assumed by \citet{Fan2006a}. 
It is also worth noting that our values for $\Gamma_{-12}$ assuming a uniform background are less than those for a non-uniform background, in agreement with the findings of previous authors [\citet{Gnedin2002,Meiksin2004, Croft2004} and \citet{Bolton2007a}].
In our model for a fluctuating IBG, high ionization rates are biased to over-dense regions and the transmissivity of rare voids remains largely unaffected by increasing the level of the IBG.
In contrast, raising the level of a uniform IBG changes the transmissivity of all regions [see \citet{Bolton2007a} and references therein].
On the other hand, the contribution of low density regions to the aggregate \Lya~absorption observed along a LOS is relatively minor, irrespective of the IBG.

\subsection{The escape fraction of ionizing photons}
\label{escape_fraction_subsection}

\begin{figure}
\vbox{\centerline{
\epsfig{file=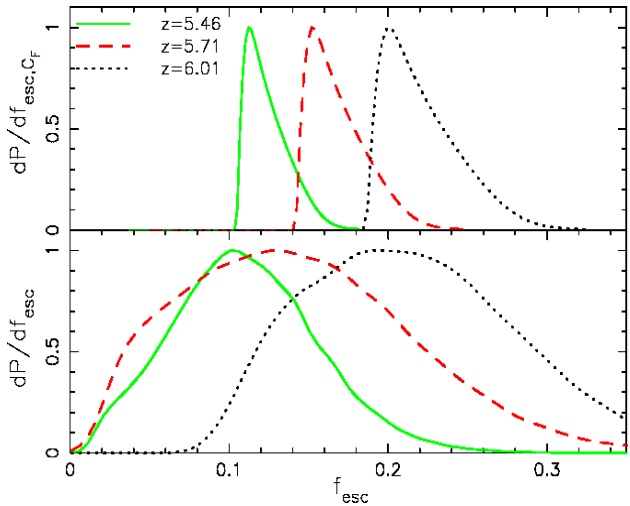,width=8.25truecm,height=6.6truecm}}
  \caption{\label{dPdfesc_NoS_figure} {
The likelihood distribution ${\frac{dP}{d\fesc}}$. 
We have adjusted the intensity of the IBG such that the model values for the mean transmission match the observed values. 
At $z=5.5,5.7,6.0$, these mean transmission values are ${\Fbareff}=0.079, 0.047, 0.006$, respectively. 
The IBG model is defined by assumed MFPs of $\lambdaMFP=28.7, 18.9, 4.2$~Mpc/$h$ respectively.
The upper panel shows the distributions of $\fesc$ corresponding to the mean transmission.
The lower panel shows the distributions of $\fesc$ after integration over the uncertainty in the mean transmission.
} }}
\end{figure}

In our model, we adjust the intensity of the IBG (via the parameter $\fesc$) in order to reproduce a given effective transmission ($F$).
Following this constraint of the escape fraction which corresponds to a particular $F$, we calculate the product ${\mathcal C}_{F}=\fstar\fesc$.
Our previous constraint of the SFR via the parameters $\fstar$~and~$\tdc$ (see \S~\ref{ionizing_background_subsection}) describes the distribution $\frac{dP}{d{\tdc d\fstar}}$.
From this distribution we then form the distribution $\frac{dP}{d{\fstar}}$ by marginalizing over $\tdc$.
Given $\frac{dP}{d{\fstar}}$, we compute the probability distribution of $\fesc$ for a particular transmission ($\frac{dP}{d{\fesc}}\bigr|_{{\mathcal C}_{F}}$) using

\begin{equation}
\label{CdPdfesc_equation}
{\frac{dP}{d{\fesc}}\biggr\vert}_{{\mathcal C}_F}=\frac{dP}{d{\fstar}} \left|\frac{d{\fstar}}{d{\fesc}}\right|_{{\mathcal C}_F} = \frac{dP}{d{\fstar}} \frac{\mathcal C_F}{{{\fesc}^{\sss 2}}}~~. 
\end{equation}

\noindent
The upper panel in Figure~\ref{dPdfesc_NoS_figure} shows the probability distributions of $\frac{dP}{d{\fesc}}\bigr|_{{\mathcal C}_{F}}$ at $z=5.5,5.7,6.0$ assuming IBGs in which the MFPs of ionizing photons are, $\lambdaMFP=28.7, 18.9, 4.2$~Mpc/$h$.
In this figure ${{\mathcal C}_F}$ corresponds to the observed mean effective transmission (i.e $F=\Fbareff$).
Our analysis suggests $\fesc \sim 10-25\%$ over the redshift range $z\sim 5.5-6$.
Theses distributions represent the statistical uncertainty in $\fesc$ owing to uncertainty in the modelling of the LF.

The lower panel in Figure~\ref{dPdfesc_NoS_figure} shows the probability distributions of $\fesc$ including the additional uncertainty owing to measurement error in the transmission.
To find this distribution,

\begin{equation}
\label{dPdfesc_equation}
\frac{dP}{d{\fesc}}\propto \int d{{\mathcal C}_F} 
{\frac{dP}{d{\fesc}}\biggr\vert}_{{\mathcal C}_F}
\frac{dP}{dF} \frac{dF}{d{{\mathcal C}_F}}~~, 
\end{equation}

\noindent
we assume the observed transmission to follow a Gaussian distribution ($\frac{dP}{dF}$) with values for the mean and variance that are given in \S~\ref{observed_transmissivity_section}.
The distributions, ${\frac{dP}{d{\fesc}}\biggr\vert}_{{\mathcal C}_F}$, can be evaluated from equation~(\ref{CdPdfesc_equation}), and ${\frac{dF}{d{{{\mathcal C}_F}}}}$ is computed from the simulations.
The uncertainty in transmission dominates the uncertainty in estimates of $\fesc$.
We find a statistical relative error on the value of $\fesc$ that is of order $\sim 50\%$, which is larger than the difference between the most likely values at each redshift.

\begin{figure}
\vbox{\centerline{
\epsfig{file=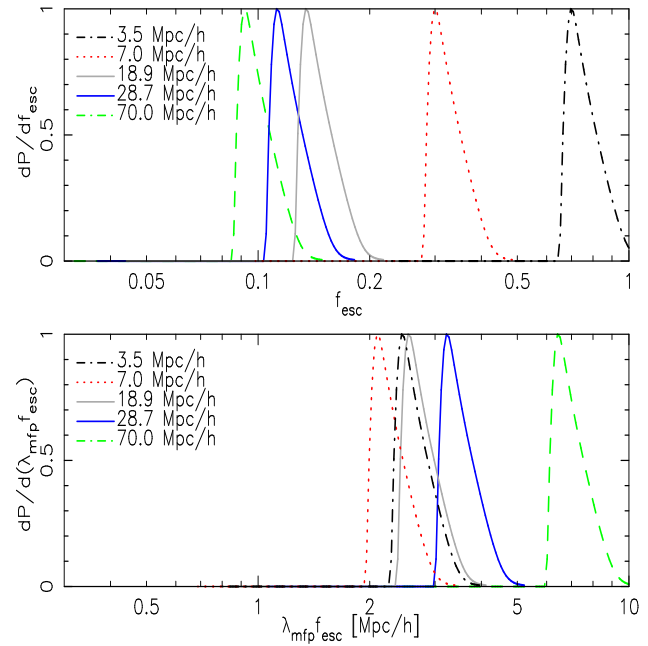,width=8.3truecm,height=8.5truecm}}
  \caption{\label{dPdfesc_NoS_panels_figure} {
Likelihood distributions for $\fesc$ and $\left(\lambda\fesc\right)$ corresponding to the observed mean transmission at $z=5.5$ for IBGs defined by MFPs $3.5, 7.0, 18.9, 28.7, 70.0$ Mpc/$h$.
We have adjusted the intensity of the IBG such that the model values for the mean transmission match the observed value. 
At $z=5.5$, the observed mean transmission is ${\Fbareff}=0.079$. 
In the upper panel the maximum likelihood occurs for $\fesc= 70, 30,13,11,9 \%$ respectively. 
In the lower panel the maximum likelihood occurs for $\lambdaMFP\fesc= 2.4,2.1, 2.5,3.2,6.3$  Mpc/$h$ respectively. 
  }  
}}
\end{figure}

The value of the MFP warrants further discussion.
The MFPs assumed in our model are based on those derived in \citet{Fan2006a} which are generally lower (by a factor of $\gtrsim 2$) than those suggested by \citet{Bolton2007a}.
The MFP estimates of \citet{Bolton2007a} assume that photons propagate freely in regions below some critical over-density which may be determined from the ionization rate and the column density ($N_{\HI}$) of self-shielded neutral clouds.
In contrast, \citet{Fan2006a} estimate the MFP considering the distribution of the neutral fraction over all over-densities in the IGM.
Although our estimates of the ionization rate are in agreement with \citet{Bolton2007a}, we choose to adopt lower values for the MFPs for two reasons.
Firstly, the MFP of ionizing photons may be over-estimated by a factor of $\sim 2$ by neglecting {\HI} absorbers between Lyman-limit systems \citep{Miralda2003,Furlanetto2005}. 
Secondly, as the MFP is very uncertain, adopting the lower values is conservative with respect to the photon budget for reionization as we will discuss in \S~\ref{small_escape_fraction_models_section}.

We note that uncertainty in the MFP represents a large source of uncertainty in our estimates of the escape fraction, and we therefore next investigate the sensitivity of the results to the MFP. 
In Figure~\ref{dPdfesc_NoS_panels_figure} we plot the distribution of the escape fraction (upper panel) corresponding to the observed mean transmission at $z=5.5$, assuming several MFPs ($3.5, 7.0, 18.9, 28.7, 70.0$ Mpc/$h$).
An increased MFP leads to a decrease in the escape fraction required to reproduce the same IBG.
For the shortest (longest) MFP considered [$\lambda=3.5 (70.0)$ Mpc/$h$] the most likely value for the escape fraction is ${\fesc}\sim 10.2\% (1.3\%)$.  
This range represents a factor of $\sim 10$ variation in the escape fraction given a range of a factor of $\sim 20$ in MFP.
In the lower panel we plot the distribution of ${\frac{dP}{d\left(\lambda\fesc\right)}}$, where 

\begin{equation}
\label{dPdfesclambda_equation}
\frac{dP}{d({\lambdaMFP\fesc})}=\frac{dP}{d{\fstar}}\frac{{\mathcal{C}_F}}{{\lambdaMFP\fesc^2}}~~,
\end{equation}
\noindent
where ${{\mathcal C}_F}$ corresponds to the mean observed transmission.
This demonstrates the level of degeneracy of the product $\lambdaMFP\fesc$.
For the shortest (longest) MFP considered [$\lambdaMFP=3.5~(70.0)$ Mpc/$h$] the most likely value in the product $\lambdaMFP{\fesc}$ is $\sim 2.4~(6.3)$ Mpc/$h$.  
This represents a factor of less than $3$. 
Thus we can summarize our results with the relation $\fesc(\lambdaMFP)\sim K_{\lambda}\left(\frac{\lambdaMFP}{{\mathrm{Mpc}/h}}\right)^{-1}$, where $2\lesssim K_\lambda \lesssim 6$.
Excluding our estimate of $\fesc$ at the longest MFP this range for $K_\lambda$ narrows to $2\lesssim K_\lambda \lesssim 3$.

\subsection{Further discussion}
\label{further_discussion_subsection}

Before proceeding it is illustrative to compare estimates of the escape fraction which are derived by \citep{Bolton2007a}.
\citet{Bolton2007a} find the escape fraction to be $20-30\%$ by comparing the constrained ionization rate to the global emissivity, which they compute by integrating the observed UV luminosity function down to a minimum galaxy brightness of $M_{1350}= -18$.
They also consider quasars to contribute to the global emissivity, however this contribution is less than $\sim 10\%$ and not considered further here.
In our model we include sources down to a lower luminosity ($M_{1350}\sim -16$) than used in \citet{Bolton2007a}, which corresponds to a factor of $\sim2$ difference in the integrated emissivity.
However this factor is partially cancelled by including the revised galaxy LF of \citet{Bouwens2007}, which leads to a decrease in the integrated emissivity of a factor of $\sim 2$.
In addition, the SED assumed in our study is $\sim 35\%$ harder [\citet{Bolton2007a}  use $\Gamma_{-12}=2.1{\jtw}$].
Finally, the definition of the escape fraction used in \citet{Bolton2007a} is equivalent to the relative escape fraction which is frequently quoted in observational studies (see \S~\ref{introduction_section}). 
This is because they infer the escape fraction of ionizing photons having assumed an emissivity which is normalized by integrating over the observed LF of LBG galaxies.
Following Bouwens et al. (2007), this emissivity should be corrected by a factor of $1.5$ to describe the intrinsic emission of galaxies (at $1350\AA$) at $z\sim6$.
An important point of difference between our work and \citet{Bolton2007a} is that the identification of halos as the sources allows the escape fraction to be connected with halo mass as well as luminosity, rather than luminosity alone.

The above corrections should be considered in order to quantitatively compare our estimate of $\fesc$ to that of \citet{Bolton2007a}.
Following these corrections the result of \citet{Bolton2007a} becomes $\fesc\sim 15-25\%$, which is consistent with our estimates. 
Although both studies are normalized to the observed LF, a significant distinction in our model is that the IBG is produced by discrete sources and so spatially non-uniform.
However, the consistency of our results with \cite{Bolton2007a} suggests that distinction between a global escape fraction derived by comparing to the integrated emissivity, and a globally averaged $\fesc$ which is apriori attributed to individual galaxies, is not significant.

Finally, we note an additional source of uncertainty in our model which is associated with the assumed isotropic escape of ionizing radiation from galaxies.
Previous studies \citep{Fujita2003,Razoumov2006,Gnedin2007a} have shown that ionizing radiation escapes from host galaxies through ionized tunnels in the ISM created by radiative feedback.
This may account for the range of values derived for $\fesc$ from observation of galaxies along different sight-lines.
In our model, this anisotropy would increase the fluctuations in the IBG. 
Based on our finding that the impact of a non-uniform IBG on the global value of $\fesc$ is not significant, we expect that the uncertainty introduced to the value of $\fesc$ by the assumption of an isotropic escape fraction in our model is not significant.


\section{Dwarf galaxy suppression}
\label{galaxy_suppression_section}

In the previous section we considered an IBG in which we included all halos down to the mass resolution of the simulation ($2\times10^9M_{\odot}$). 
In our best fit model this mass corresponds to a luminosity of $M_{1350} \sim -16$. 
In this section we investigate the impact on the inferred escape fraction of allowing the suppression of dwarf galaxies.
The reionization of the IGM elevates the Jeans mass and so increases the required minimum virial temperature a DM halo must have in order for gas to accrete and form stars. 
This minimum virial temperature could be as large as $\sim2.5\times10^5$ K which corresponds to a minimum halo mass,   $M_{vir}\sim 10^{10}M_\odot$ \citep{Efstathiou1992,Thoul1996,Dijkstra2004b}.
The suppression of small galaxies will modify both the topology and the level of the IBG and hence also the derived estimate of the escape fraction.
We repeat our analysis of the galaxy LF, limiting the mass of halos which are considered as potential sources of UV radiation to $M>10^{\sss 10}~M_{\odot}$.

\begin{figure}
\vbox{\centerline{
\epsfig{file=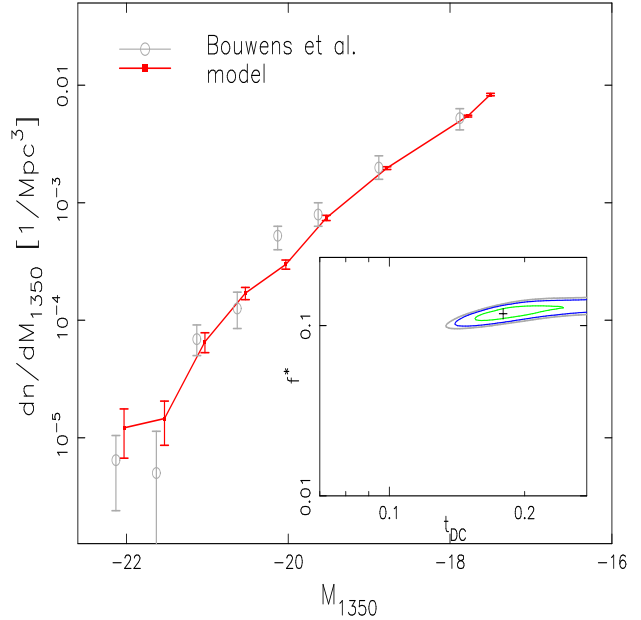,width=8.2truecm}}
\caption{\label{bestfitLF_S}
{
The best fit model LF with parameter values of ${\tdc}=0.18~({\ts}=2.5\times10^8$~yr) and $f^*=0.12$, as constrained by a maximum likelihood analysis.
Inset is is a plot of the likelihood distribution in which the contours represent the $1,2,3-\sigma$ confidence levels with a cross to mark the point of maximum likelihood. 
For display purposes a line joins the model points (solid). 
The error bars on the model reflect the number of galaxies in each luminosity bin as measured from our simulation.
The open circles represent the data from the observed UV galaxy LF \citep{Bouwens2007}. 
The model points are shifted slightly to the right of the observational data for clarity.
}}
}
\end{figure}

Repeating our analysis of section~\ref{ionizing_background_subsection}, Figure~{\ref{bestfitLF_S}} shows the best fit LF for the suppression of galaxies with halo mass less than $10^{10}\Msun$.
For display purposes a line joins the model points (solid) which are the centers of the constructed luminosity bins.
At the faint end of the LF we have constructed an additional bin as we are able to resolve galaxies down to a lower luminosity in our simulation than are currently observed.
The error bars on the model reflect the number of galaxies in each luminosity bin.
The open circles represent the data from the observed UV galaxy LF \citep{Bouwens2007}. 
The model points are shifted slightly to the right of the observational data for clarity only. 
Inset in Figure~{\ref{bestfitLF_S}} is a plot of contours representing the $1,2,3-\sigma$ confidence levels on the parameters $\fstar$ and $\tdc$.
A cross marks the point of maximum likelihood, with values of ${\tdc}=0.18~({\ts}=2.5\times10^8$~yr) and $f^*=0.12$. 
These values are not significantly different from those previously derived for the inclusion of galaxies with halo mass down to $10^9\Msun$ (\S~\ref{ionizing_background_subsection}).
For the suppression of galaxies with halo mass less than $10^{10}\Msun$, the minimum mass halo which contributes to the IBG corresponds to a specific luminosity of $L\sim 4.3\times10^{27}$ ergs/sec/Hz and a magnitude of $M_{\scs{1350}}\sim-17.5$.
This luminosity is smaller than the minimum luminosity included in \citet{Bolton2007a}.

\begin{figure}
\vbox{\centerline{
\epsfig{file=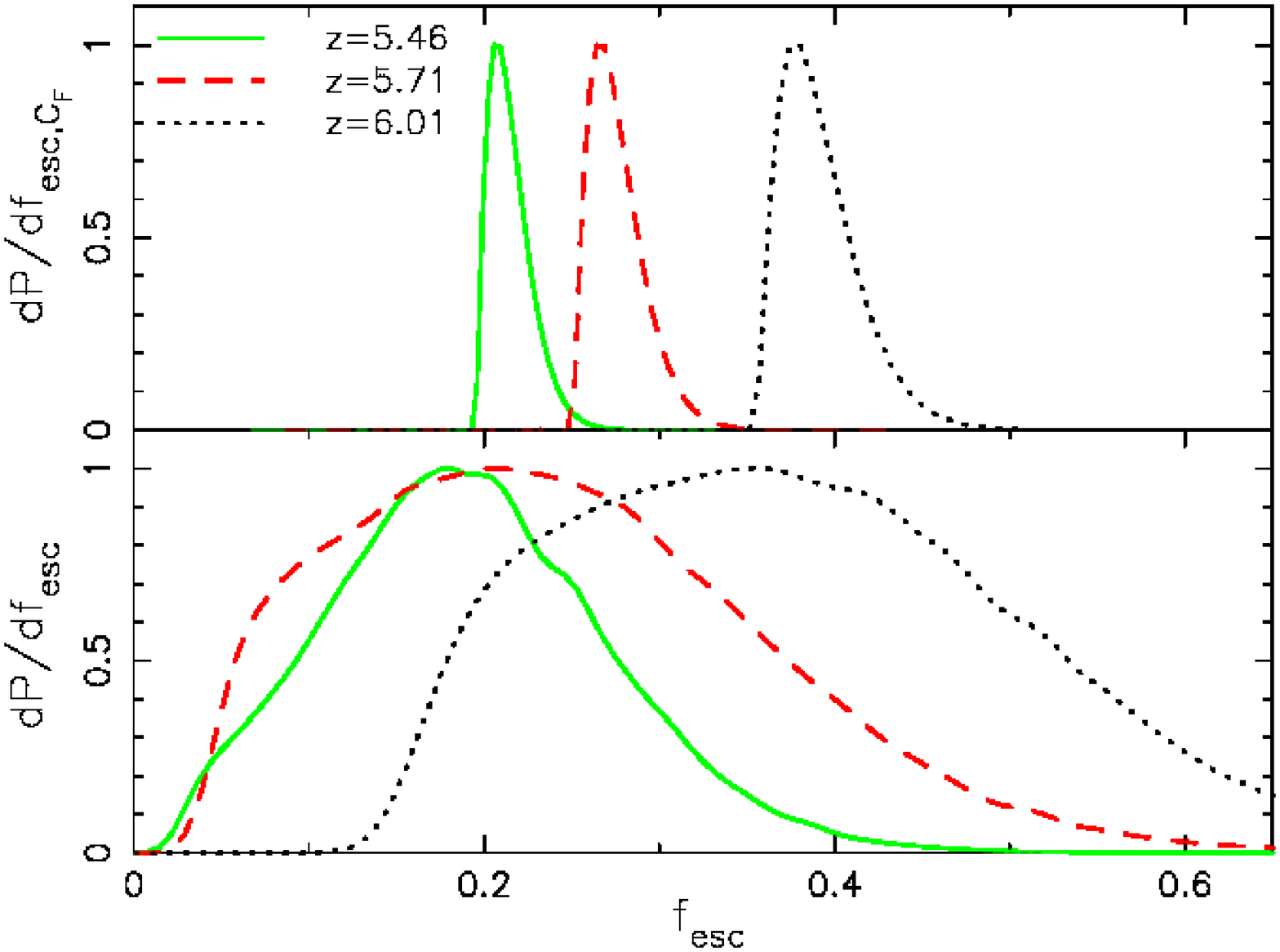,width=8.25truecm,height=6.5truecm}}
  \caption{\label{dPdfesc_figure_S} {
The likelihood distribution ${\frac{dP}{d\fesc}}$. 
We have adjusted the intensity of the IBG such that the model values for the mean transmission match the observed values. 
At $z=5.5,5.7,6.0$, these mean transmission values are ${\Fbareff}=0.079, 0.047, 0.006$, respectively. 
The IBG model is defined by assumed MFPs of $\lambdaMFP=28.7, 18.9, 4.2$~Mpc/$h$ respectively.
The upper panel shows the distributions of $\fesc$ corresponding to the mean transmission.
The lower panel shows the distributions of $\fesc$ after integration over the uncertainty in the mean transmission.
  }  
}}
\end{figure}

Finally, we compute the likelihood distributions for the escape fraction, analogous to those shown Figure~\ref{dPdfesc_NoS_figure}.
In the upper panel of Figure~\ref{dPdfesc_figure_S} we show the probability distributions of $\fesc$ at our fiducial redshifts and MFPs, where we have constrained the transmission to the mean observed transmission.
This yields values for ${\fesc}$ between $20\%$ and $40\%$, which are larger than those found in Figure~\ref{dPdfesc_NoS_figure}, where we included sources down to a halo mass of $\sim 10^9\Msun$.
The distributions in the upper panel of Figure~\ref{dPdfesc_figure_S} represent uncertainty in $\fesc$ owing to uncertainty in the modelling of the LF.
The lower panel in Figure~\ref{dPdfesc_figure_S} shows the probability distributions of $\fesc$ for the same redshifts and MFPs as the lower panel, however here we have included the additional uncertainty introduced by the distribution of optical depth measurements.
As before, we assume the observed transmission to follow a Gaussian distribution ($\frac{dP}{dF}$) with values for the mean and variance that are given in \S~\ref{observed_transmissivity_section}.

\section{Reionization by massive galaxies}
\label{reionization_massive_galaxies_section}

\begin{figure}
\vbox{\centerline{
\epsfig{file=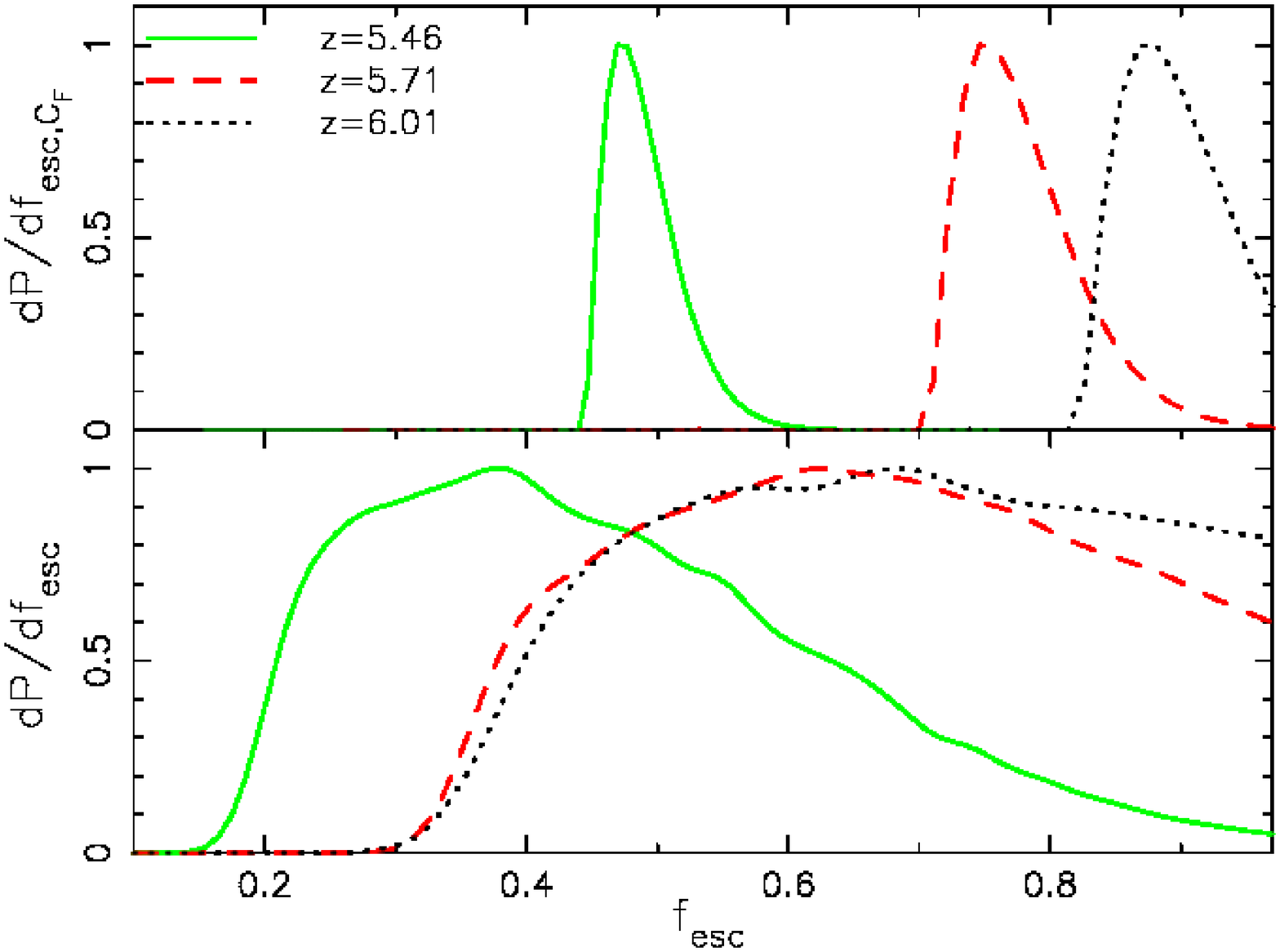,width=8.25truecm,height=6.5truecm}}
  \caption{\label{dPdfesc_figure_S11} {
The likelihood distribution ${\frac{dP}{d\fesc}}$. 
We have adjusted the intensity of the IBG such that the model values for the mean transmission match the observed values. 
At $z=5.5,5.7,6.0$, these mean transmission values are ${\Fbareff}=0.079, 0.047, 0.006$, respectively. 
The IBG model is defined by assumed MFPs of $\lambdaMFP=28.7, 18.9, 4.2$~Mpc/$h$ respectively.
The upper panel shows the distributions of $\fesc$ corresponding to the mean transmission.
The lower panel shows the distributions of $\fesc$ after integration over the uncertainty in the mean transmission.
  }  
}}
\end{figure}

In a recent paper, \citet{Gnedin2007a} suggested that the escape fraction of ionizing photons is a few percent for massive galaxies, but is negligibly small for galaxies which reside in halos of mass $M\lesssim 5\times10^{10} M_{\odot}$.
We investigate the implications for the escape fraction inferred from {\Lya} transmission data by excluding halos of mass $M<5\times 10^{10} M_{\odot}$.
The suggestion that ionizing photons do not escape galaxies with halo mass less than $5\times10^{10} M_{\odot}$ does not imply that photons above the Lyman-limit (i.e. $1350 \AA$) are similarly affected.
Therefore we assume the values for $\fstar$ and $\tdc$ from \S~\ref{galaxy_suppression_section} to compute the IBG in this case. 
In this case, the minimum halo mass corresponds to a minimum specific luminosity of $L\sim 2.2\times10^{28}$ ergs/sec/Hz and a magnitude of $M_{\scs{1350}}\sim-19.3$, which is brighter than the minimum magnitude considered by \citet{Bolton2007a}.

Again we compute the likelihood distributions for the escape fraction.
In the upper panel of Figure~\ref{dPdfesc_figure_S11} we show the probability distributions of $\fesc$ at our fiducial redshifts and MFPs, where we have constrained the transmission to the mean observed transmission.
This yields values for ${\fesc}$ between $46\%$ and $86\%$, which are larger again than those found in Figure~\ref{dPdfesc_NoS_figure}, where we included sources down to a halo mass of $\sim 10^9\Msun$.
As previously computed in \S~\ref{escape_fraction_subsection}, the lower panel in Figure~\ref{dPdfesc_figure_S11} shows the distributions of $\fesc$ in which we have included the uncertainty introduced by the distribution of optical depth measurements at each redshift.
These results imply that reionization by massive galaxies alone, with escape fractions of order a few percent \citep{Gnedin2007a} is not consistent with the measured ionization rate at $z\sim 5-6$.

\begin{figure}
\vbox{\centerline{
\epsfig{file=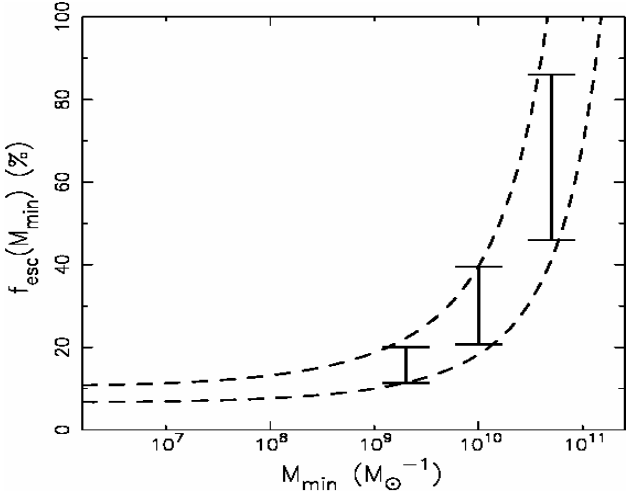,width=8.25truecm,height=6.5truecm}}
  \caption{\label{MminVfesc_figure} {
Extrapolation of estimates of $\fesc$ as a function of minimum mass from numerical simulations.
Dashed lines show analytic fits to the numerical data for both the maximum and minimum values found for the normalization parameter $A$ (see text).
The numerical estimates are shown including the range for the median value of $\fesc$ predicted in our simulation at the three redshifts modelled.
The analytic fit suggests that the value for $\fesc$ is greater than $100\%$ for minimum masses, $M_{\rm{min}} \gtrsim 10^{12}$, and approaches a constant value of $\sim5-10\%$ for minimum masses near the hydrogen cooling threshold.
Figure~\ref{MminVfesc_figure} indicates that the escape fraction from high redshift galaxies must be $\gtrsim 5\%$ irrespective of the minimum mass.
  }  
}}
\end{figure}

\section{Extrapolation of numerical results to other galaxy masses}
\label{extrapolation_section}

In Figure~\ref{MminVfesc_figure} we summarize our estimates of the escape fraction for the three minimum mass scenarios we have considered.
The numerical estimates are shown including the range for the median value of $\fesc$ predicted in our simulation at the three redshifts modelled.
For comparison, the dashed lines represent simple analytic fits to describe the relationship between the minimum mass of halos which contribute to the IBG and the escape fraction.
To estimate this relationship we assume that the product of the escape fraction and the star-formation rate is constant and that the star-formation rate can be approximated as being proportional to the time derivative of the collapsed mass faction, $dF_{\rm{col}}/dt$.
The quantity $dF_{\rm{col}}/dt$ is computed analytically as a function of halo mass and redshift using the prescription of \citep{Sheth}.
Thus the escape fraction can be expressed as 

\begin{equation}
\label{dFdt_equation}
\fesc(M_{\rm{min}})=  A(M_{\rm{min}}) \left[\frac{dF_{\rm{col}}}{dt}(M_{\rm{min}})\right]^{-1}~~,
\end{equation}

\noindent
where $M_{\rm{min}}$ is the minimum halo mass of galaxies which contribute to the IBG, and $A$ is a constant.

We normalize the relation between the escape fraction and the star-formation rate using our estimates of $\fesc$ from the simulations (at $z=5.5, 5.7, 6.0$), in which the minimum contributing halo masses ($M=2\times10^{9},10^{10},5\times10^{10}$\msun) are adopted to compute $dF/dt$.
The analytic time derivative of the collapsed mass faction ($dF_{\rm{col}}/dt$) does not perfectly describe the SFR from the simulation. 
The value of $A$ is therefore mass and redshift dependent and the relative variation in this normalization parameter is found to be $7\%~(11\%)$ for minimum mass halos of $M=2\times10^{9}-5\times10^{10}$.
In Figure~\ref{MminVfesc_figure} we plot $\fesc(M_{\rm{min}})$ (dashed lines) assuming both the maximum and minimum values found for $A$.

Figure~\ref{MminVfesc_figure} shows that as the minimum mass is increased, the global escape fraction increases to compensate for the lost emissivity.
The analytic fit suggests that the value for $\fesc$ is greater than $100\%$ for minimum masses, $M_{\rm{min}} \gtrsim 10^{12}$, indicating an upper limit to $M_{\rm{min}}$.
The values of $\fesc(M_{\rm{min}})$ can be extrapolated to masses below the mass resolution limit of the simulation ($2\times10^9$\msun) as shown in Figure~\ref{MminVfesc_figure}.
The escape fraction approaches a constant value of $\sim5-10\%$ for minimum masses near the hydrogen cooling threshold.
These escape fractions are comparable to observed estimates for individual galaxies at lower redshift.
Interestingly, Figure~\ref{MminVfesc_figure} indicates that the escape fraction from high redshift galaxies must be $\gtrsim 5\%$ irrespective of the minimum mass.

\section{Models of the reionization history}
\label{small_escape_fraction_models_section}

\begin{figure*}
\includegraphics[width=15cm]{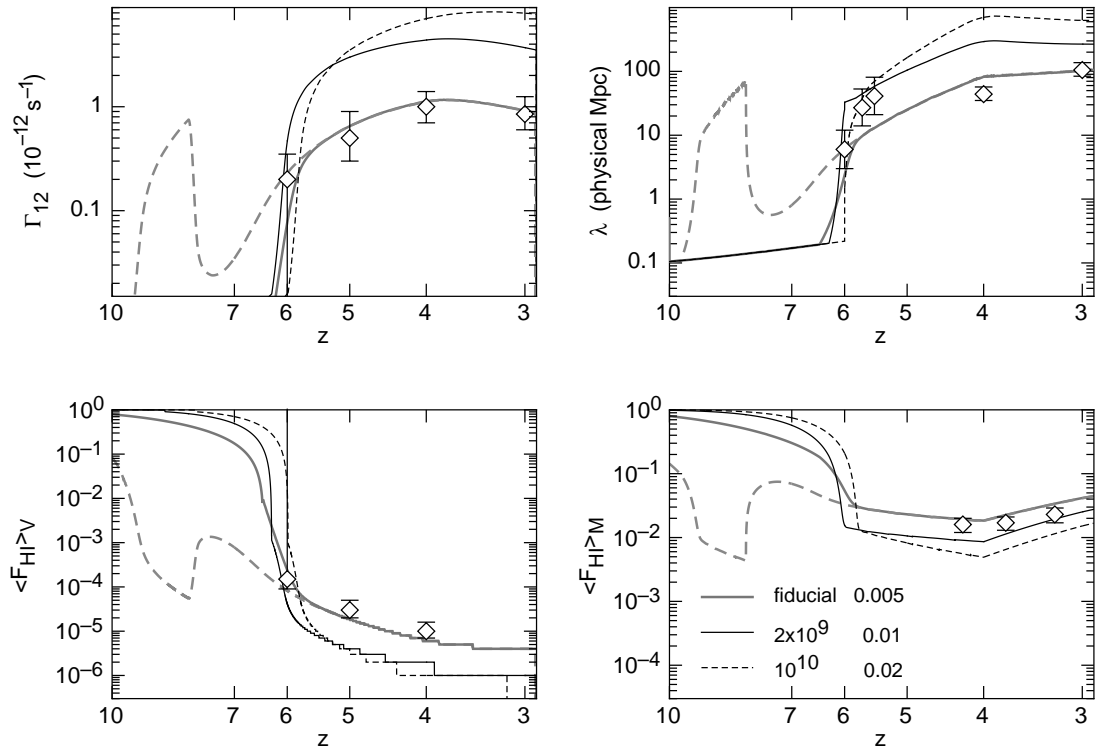} 
\caption{Models for the reionization of the IGM and the subsequent
post-overlap evolution of the ionizing radiation field. 
In each panel the case shown corresponds to a value for the critical over-density prior to the overlap epoch of $\Delta_{\rm c}=5$. 
{\em Upper Left Panel:} The ionization rate as a function of redshift. The observational points are from \citet{Bolton2007a}. 
{\em Lower Left and Right Panels:} The volume and mass
averaged fractions of neutral gas in the universe. The observational
points for the volume averaged neutral fraction are from \citep{Bolton2007a}, while the observed mass-fractions are from the damped {\Lya} measurements of Prochaska et al.~(2005).  {\em Upper Right Panel:} The mean-free-path for ionizing
photons. The data points
are based on Storrie-Lombardi et al.~(1994). The models are labelled by the assumed values of $M_{\rm esc}$ and $\fstaresc$. }
\label{variance_history}
\end{figure*} 

The {\Lya} absorption probes the instantaneous emissivity of the galaxy population, and hence allows the constraints on the escape fraction presented thus far.
However, the redshift at which reionization occurs is dependent on the cumulative star-formation.
In the remainder of this paper we discuss constraints on the reionization history.

\subsection{Semi-analytic model for reionization}
\label{models}

In this section we use a semi-analytic model to study the implications of the implied values of escape fraction on the reionization history of the IGM.
\citet{Miralda2000} presented a formalism which allows the calculation of an effective recombination rate in an inhomogeneous universe by assuming a maximum over-density ($\Delta_{\rm c}$) penetrated by ionizing photons within \HII~regions. 
Their model assumes that reionization progresses rapidly through islands of lower density prior to the overlap of individual cosmological ionized regions. 
Following the overlap epoch, the remaining regions of high density are gradually ionized. 
It is therefore hypothesized that at any time, regions with gas below some critical over-density $\Delta_{\rm i}\equiv {\rho_{i}}/{\langle\rho\rangle}$ are highly ionized while regions of higher density are not. 
In what follows, we draw primarily from their prescription and refer the reader to the original paper for a detailed discussion of its motivations and assumptions.  
\citet{Wyithe2003} employed this prescription within a semi-analytic model of reionization. 
This model was extended by Srbinovsky \& Wyithe~(2007) and by \citet{Bolton2007a}.
We refer the reader to those papers for a full description.

Within the formalism of Miralda-Escude et al.~(2000) we describe the post-overlap evolution of the IGM by computing the evolution of the fraction of mass in regions with over-density below $\Delta_{\rm i}$,

\begin{equation}
F_{\rm M}(\Delta_{\rm i})=\int_{0}^{\Delta_{\rm i}}d\Delta P_{\rm
V}(\Delta)\Delta~~,
\end{equation}

\noindent
where $P_{\rm V}(\Delta)$ is the volume weighted probability distribution for
$\Delta$. 
Miralda-Escude et al.~(2000) quote a fitting function which provides a good fit to the volume weighted probability distribution for the baryon density in cosmological hydrodynamical simulations. 
This probability distribution remains a reasonable description at high redshift when confronted with a more modern cosmology and updated simulations, although an analytical approximation for the high density tail of the distribution remains necessary as a best guess at correcting for numerical resolution \citep{Bolton2007a}.

Prior to the overlap epoch, a constant value $\Delta_i=\Delta_{\rm c}$ is assumed, while in the post overlap era our model for the reionization history computes the evolution of $\Delta_i$. 
We compute evolution of the quantity $Q_{\rm i}$ which is defined to be the volume filling factor within which all matter at densities below $\Delta_{\rm c}$ has been ionized. 
Within our formalism, the epoch of overlap is precisely defined as the time when $Q_{\rm i}$ reaches unity. 
However, we have only a single equation to describe the evolution of two independent quantities $Q_{\rm i}$ and $F_{\rm M}$ (or equivalently $\Delta_{\rm c}$). 
The relative growth of these depends on the luminosity function and spatial distribution of the sources. 
We assume $\Delta_{\rm c}$ to be constant with redshift before the overlap epoch and compute results for models with values of $\Delta_{\rm c}=5$. 
Within the formalism of Miralda-Escude et al.~(2000), a maximum over-density of $\Delta_{\rm c}=5$ corresponds approximately to the mean separation between galaxies at high redshift.
For a large range of values, the effect of varying $\Delta_{\rm c}$ is smaller than other uncertainties in the problem \citep{WyitheandBolton2007}.

Our approach is to compute a reionization history given a particular value of $\Delta_{\rm c}$, combined with assumed values for the efficiency of star-formation and the fraction of ionizing photons that escape from galaxies. 
With this history in place we then compute the evolution of the background radiation field due to these same sources.  
After the overlap epoch, ionizing photons will experience attenuation due to residual over-dense pockets of HI gas.
We use the description of Miralda-Escude et al.~(2000) to estimate the ionizing photon mean-free-path, and subsequently derive the attenuation of ionizing photons. 
We then compute the flux at the {\Lylim} in the IGM due to sources immediate to each epoch, in addition to redshifted contributions from earlier epochs. 

As in earlier sections we assume an IBG dominated by galaxies and the spectral energy distribution (SED) of population-II star forming galaxies, using the model presented in Leitherer et al.~(1999). 
The star-formation rate per unit volume is computed based on the collapsed fraction obtained from the extended \citet{Press1974} model \citep{Bond1991} in halos above the minimum halo mass for star formation, together with an assumed star formation efficiency.

In a cold neutral IGM beyond the redshift of reionization, the collapsed fraction should be computed for halos of sufficient mass to initiate star formation. 
The critical virial temperature is set by the temperature ($T_{{\mathrm{N}}}\sim 10^4$ K) above which efficient atomic hydrogen cooling promotes star formation. 
Following the reionization of a region, the Jeans mass in the heated IGM limits accretion to halos above $T_{{\mathrm{I}}}\sim10^5$ K (Efstathiou~1992; Thoul \& Weinberg~1996; Dijkstra et al.~2004b).
As described in the introduction, only a fraction of ionizing photons produced by stars enter the IGM. Therefore an additional factor of $f_{\mathrm{esc}}$ (the escape fraction) must be included when computing the emissivity of galaxies. 
In our fiducial model we assume this escape fraction to be independent of mass. 
We define a parameter
$\fstaresc\equiv \fstar \fesc$.

Figure~\ref{variance_history} shows our fiducial model for the reionization of the IGM and the subsequent post-overlap evolution of the ionizing radiation field, 
assuming $\fstaresc=0.005$ ($\fesc\sim5\%$).  
In the top left panel of Figure~\ref{variance_history} we show the evolution of the ionization rate. 
The observational points are from the simulations of \citet{Bolton2007a}, which at the redshifts shown are based on observations \citep{Songaila2004,Fan2006a} and simulated data \citep{Schaye2003} of the \Lya~opacity of the IGM.
The assumed value of $\fstaresc=0.005$ provides the best fit to the data, and is consistent with modelling of the LF ($\fstar\sim0.1$) as well as extrapolation of the escape fraction to minimum masses below $10^9$~\msun ($\fesc\sim5\%$).

In the lower-left and lower-right panels we plot the corresponding volume  and mass (upper curves) averaged fractions of neutral gas in the universe. 
The observational points for the volume averaged neutral fraction are from \citet{Bolton2007a}, while the observed mass-fractions are from the damped {\Lya} measurements of \citet{Prochaska2005}, and therefore represent lower limits on the total HI content of the IGM. 
Both curves for the mass fraction and the volume fraction show excellent agreement with these observed quantities.
This is despite their differing by 3 orders of magnitude and indicates the applicability of the model over a wide density range.
In the upper-right panel we plot the evolution of the ionizing photon mean-free-path. 
The data points at  $z>5$ are based on \citet{Storrie1994}. 
At $z>5$ the values of the MFP adopted in our simulations are shown for comparison.
Again the model is in good agreement with the available observations and the estimated MFPs.
The observed values for the MFP obtained by \citet{Storrie1994} are found from the number density of Lyman-limit systems (which represents an upper limit, see discussion in \S~\ref{escape_fraction_subsection}) and is independent of the Ly$\alpha$ forest absorption derived quantities of ionization rate and volume averaged neutral fraction, as well as being independent of the HI mass-density measurements.
Our simple model therefore simultaneously reproduces the evolution of three independent measured quantities. 

The agreement of this model with observations is an important consistency check of our parameter estimates in previous sections. 
Taken together, our results for the escape fraction and star-formation efficiency based on our numerical simulations combined with the model reionization histories suggest that the observed transmission and luminosity function are compatible with reionization at $z\sim6$.

\subsection{Comparison of the escape fraction in numerical and semi analytic models}

Our numerical simulations are only able to resolve galaxies down to a halo mass of $M=2\times10^9M_\odot$, while in our fiducial reionization model the bulk of ionizing photons prior to reionization come from lower mass galaxies ($M\gtrsim10^{8}M_\odot$). 
We therefore also use our semi-analytic history model to investigate the implications of excluding these lower mass galaxies, as well as galaxies with halo mass below  $10^{10}$\msun (see \S~\ref{galaxy_suppression_section}).
We adopt values for the escape fraction of $\fstaresc=0.01$ and $\fstaresc=0.02$ respectively, which are consistent with our numerical results in the previous sections.
The results of these models are plotted in Figure~\ref{variance_history} (thin and dashed lines).
These histories are consistent with reionization at $z\sim6$, which directly constrains the the result that suggests that the observed transmission and luminosity function are compatible with reionization at $z\sim6$.
However, these models exceed the observed ionization rate at $z\lesssim6$, owing to the rapid increase in the collapsed fraction.

In addition to the properties described in Figure~\ref{variance_history}, reionization histories are also constrained by the optical depth to Thomson scattering of CMB photons.  
The fiducial model in Figure~\ref{variance_history} has an optical depth of $\tau_{\rm es}=0.063$, while models with $M_{\rm esc}=10^9M_\odot$ and $M_{\rm esc}=10^{10}M_\odot$  have $\tau_{\rm es}=0.045$, $\tau_{\rm es}=0.041$.  
These values are smaller than the value determined from the {\em WMAP} satellite, $\tau_{\rm es}=0.084\pm0.016$ \citep{Komatsu2008}.

\subsection{Reionization histories with Massive Stars}

Before concluding, we consider the implications of a population of massive stars at high redshift.
Models which include massive population-III stars partially reionize the IGM at high redshift and result in larger values of $\tau_{\rm es}$. 
For example, in the work of \citet{Choudhury2005} and \citet{WyitheandCen2007}, models in which the reionization history shows an extended plateau of mostly ionized IGM in the range $6\la z\la10$ are able to produce values of $\tau_{\rm es}$ within the preferred range plus completion of reionization at $z\sim6$. 
This early reionization does not however make our predicted values of the escape fraction inconsistent with reionization at \zsix.

The heavy dashed line in Figure~\ref{variance_history} represents a model for the reionization history which includes population-III stars at high redshift. 
Here we assume that the value of $\fstaresc$ is not sensitive to redshift, but that 10 times the number of ionizing photons per baryon are emitted prior to a transition redshift $z_{\rm tran}$. 
Note that this modelling is intended to be illustrative only and does not include more advanced treatments of the transition redshift, such as those studied by \citet{Schneider2006}, \citet{Scannapieco2003} and Wyithe \& Cen~(2007). 
The example shown illustrates the effect of a high redshift population of massive stars and corresponds to an assumed transition redshift of $z_{\rm tran}=8$. 
Finally, the addition of population-III stars increases the value of $\tau_{\rm es}$. 
For the inclusion of population-III stars in our fiducial model, where $z_{\rm tran}=8$, we find $\tau_{\rm es}=0.095$, which is consistent with the results from the \wmap~satellite ($\tau_{\rm es}=0.084\pm0.016$).


\section{CONCLUSIONS}
\label{conclusions_section}

In this paper we have employed an $N$-body simulation to describe the density and peculiar velocity fields in a $(65.6$~Mpc/$h)^{\sss 3}$ volume of the IGM at $z=5.5,5.7,6.0$.  
The simulation also identified the locations and masses of virialized halos, with the minimum halo mass accurately resolved by the simulation being $2\times10^9M_{\odot}$.  
We associate the identified halos with galaxies and assign to them a UV luminosity using models for the SFR and the SED. 
By fitting to the observed UV galaxy LF \citep{Bouwens2007} we constrained the best fit free parameters to be ${\fstar}~=0.11$ and ${\tdc}~=0.16$, where $\fstar$ is the fraction of baryons which participate in star formation and $\tdc$ is the duty-cycle of star-bursting galaxies.  
We then constructed an ionizing background which reflects the spatial distribution of UV luminous galaxies.  
Our model made the assumption of a universal MFP at each redshift.

Given the density, velocity and ionizing radiation fields we extracted an ensemble of {\Lya} absorption spectra along 3072 sight-lines. 
From this sample we computed the mean effective transmission as a function of the fraction of ionizing photons produced that escape their host galaxy ({$\fesc$}). 
In our simulations $\fesc$ effectively adjusts the intensity of the ionizing background. 
The value of $\fesc$ was determined so that the mean effective transmission predicted by the model matched the observed value. 
Assuming halos above the mass resolution of our simulation to host galaxies, we find values for the escape fraction at $z\sim5.5-6$ of $\fesc \sim 10-25\%$
assuming values for the MFP which are suggested by \citep{Fan2006a}.
In models where we assume the formation of galaxies to be suppressed in halos with masses below $10^{10}M_{\odot}$ (possibly due to radiation feedback following reionization), we find the escape fraction to be in the range $\fesc \sim 20-40\%$. 
Our results for the escape fraction are consistent with those found in \citet{Bolton2007a} who constrain the escape fraction to be $\fesc \sim 20-30\%$, using a similar approach to the one adopted in this paper.
This is true both for scenarios in which the minimum halo mass is set by the simulation resolution ($2\times 10^{\scs 9}M_{\odot}$) and for the larger value of $10^{\scs 10} M_{\odot}$.

We also considered the scenario suggested by \citet{Gnedin2007a} in which the escape fraction of ionizing photons is negligible in halos with masses below $\sim5\times10^{10}M_\odot$.
In this case we find the escape fraction is required to be $\fesc \sim 46-86\%$ in order to reproduce the observed optical depth of the IGM.
This is inconsistent with theoretical modelling of the escape fraction by \citet{Gnedin2007a} who suggest $\fesc\sim1-3\%$.
Using the numerical results to calibrate an analytic relation between the escape fraction and minimum galaxy halo mass we extrapolate our results to a mass ($M\sim10^8$~\msun) corresponding to the hydrogen cooling threshold.
In this case we find $\fesc\sim5-10\%$, consistent with observed estimates at lower redshift.

Since the SFR is estimated at $1350\AA$, while $\fesc$ is estimated at $912\AA$, the escape fraction is degenerate with the adopted galaxy SED.
The value of the Lyman-break factor ($LB$), which is the intrinsic flux decrement measured across the Lyman break, is particularly sensitive to to the age of the stellar population. 
Assuming that the normalization of the adopted SED at wavelengths above the Lyman-limit, and the slope of the SED at wavelengths below the Lyman-limit are both correct, we can express this degeneracy with respect to LB as $\fesc=\fesc^{\sssrm{LB}=3}\left({{LB}}/3 \right)$.
In this expression $\fesc^{\sssrm{LB}=3}$ is the escape fraction which we have constrained in this paper, and $\fesc$ the value obtained for an alternate $LB$.

We use a semi-analytic model that describes the evolution of the ionized state of the post-reionization IGM to study whether the escape fraction implied by our numerical modelling is consistent with reionization at $z\gtrsim6$. 
We find that the measured ionization rates imply a history that is consistent with reionization at \zsix~for a range of these models, including those in which the minimum halo mass of galaxies contributing to the ionizing emissivity is $M\sim10^8,10^{9}$ and $10^{10}$ solar masses respectively.
However, we find that models with minimum masses above $10^9$~\msun~over-estimate the ionization rate in the post-overlap IGM.
These models imply that the Universe was reionized by low mass galaxies with an escape fraction of $\sim 5\%$.

In summary, the transmission of \Lya~photons through the IGM at $z\sim5.5-6$ implies an ionization rate that requires the escape fraction of ionizing photons to be in excess of $5\%$.
Our results for the escape fraction and star-formation efficiency based on our numerical simulations combined with the model reionization histories suggest that the observed transmission and luminosity function are compatible with reionization at $z\sim6$.

\section*{Acknowledgments}
We thank Adam Lidz for providing the $N$-body simulation and technical assistance, and also the Institute for Theory and Computation at CfA (Harvard-Smithsonian) for use of their computational facilities during this project.
We also thank an anonymous referee, whose detailed report considerably improved this paper.
This work was supported in part by the Australian Research Council. 
JAS acknowledges the support of an Australian Postgraduate Award.

\bibliographystyle{mn2e}
\bibliography{scatter}
\end{document}